\pdfoutput=1
\documentclass[preprint,aip,pof]{revtex4-1}
\usepackage[utf8]{inputenc}
\usepackage{amssymb}
\usepackage{graphicx}
\usepackage{subfigure}
\usepackage{bm}
\usepackage{amsmath}
\usepackage{threeparttable}
\usepackage{CJK}
\usepackage{hyperref}
\usepackage{color}
\usepackage{booktabs}
\usepackage{bookmark}
\usepackage{array}
\usepackage{multirow}
\usepackage{setspace}

\begin{document}
\begin{CJK*}{UTF8}{gbsn}
\title{An implicit kinetic inviscid flux for predicting continuum flows in all speed regimes}
\author{Junzhe Cao (曹竣哲)}
\email[]{caojunzhe@mail.nwpu.edu.cn}
\affiliation{School of Aeronautics, Northwestern Polytechnical University, Xi'an, Shaanxi 710072, China}
\author{Sha Liu (刘沙)}
\email[Corresponding author:]{shaliu@nwpu.edu.cn}
\affiliation{National Key Laboratory of Science and Technology on Aerodynamic Design and Research, Northwestern Polytechnical University, Xi'an, Shaanxi 710072, China}
\author{Chengwen Zhong (钟诚文)}
\email[]{zhongcw@nwpu.edu.cn}
\affiliation{National Key Laboratory of Science and Technology on Aerodynamic Design and Research, Northwestern Polytechnical University, Xi'an, Shaanxi 710072, China}

\begin{abstract}
In this study, the kinetic inviscid flux (KIF) is improved and an implicit strategy is coupled. The recently proposed KIF is a kind of inviscid flux, whose microscopic mechanism makes it good at solving shock waves, with advantages against the shock instability phenomenon. When developing the implicit KIF, a phenomenon is noticed that the kinetic flux vector splitting (KFVS) part in boundary layers not only reduces the accuracy, but seriously reduces the Courant-Friedrichs-Lewy (CFL) number as well. As a result, in this paper, a new weight is proposed about how to combine the KFVS method well with the totally thermalized transport (TTT) method. Besides admitting the using of larger CFL numbers, this new weight brings about more accurate numerical results like pressure, friction coefficient and heat flux when solving shock waves, boundary layers and complex supersonic/hypersonic flows. In order to examine the validity, accuracy and efficiency of the present method, six numerical test cases, covering the whole speed regime, are conducted, including the hypersonic viscous flow past a cylinder, the hypersonic double-cone flow, the hypersonic double-ellipsoid flow, the laminar shock boundary layer interaction, the supersonic flow around a ramp segment and the lid-driven cavity flow. The advantages of this scheme and corresponding mechanisms are to be discussed in detail.
\end{abstract}

\maketitle

\end{CJK*}

\section{\label{sec:introduction}Introduction}
In the computational fluid dynamic (CFD), the finite volume method (FVM)~\cite{jst,jiri,leveque} is one of the most popular approaches to solve flow fields numerically. Reconstruction, evolution and average are three basic steps for solving the advection equation in the classical FVM, and evolution is of vital importance. Based on the Euler equation, kinds of numerical schemes are developed for the evolution step for numerical inviscid flux, with advantages and disadvantages. For example, the Jameson-Schmidt-Turkel (JST) scheme series~\cite{jst} work well when solving subsonic and transonic flows, benefiting from the perfect use of artificial viscosity, but it is not easy to be controlled accurately in hypersonic flows~\cite{jstzj}. The flux difference splitting (FDS) schemes, like the Roe scheme series~\cite{roe} and the Harten-Lax-van Leer (HLL) scheme series~\cite{hll,hllc}, are developed from the solution of the Riemann problem, with good artificial viscosity in both subsonic and hypersonic flows, while shock instability phenomenon exists~\cite{quirk} and plenty of technologies for modification are proposed~\cite{roem,rotatedkk}. Flux vector splitting (FVS) schemes, like the Steger-Warming scheme series~\cite{sw} and the van Leer splitting scheme series~\cite{vl}, are good at dealing with the shock instability phenomenon, but they have deficiency in solving boundary layers~\cite{vl}. The advection upstream splitting method (AUSM) series separate out the pressure item and combine the FDS and the FVS together, which make success~\cite{ausm,ausmplusup}.

In another way, the flow evolution at discrete cell interfaces can be described by solving the Boltzmann equation in the gas-kinetic theory. Moreover, the numerical flux determined by the Maxwellian distribution of an equilibrium state plays the same role of those from the Euler equation. The existing schemes based on the gas-kinetic theory often have different properties because of their different ways in using the microscopic information. For example, the equilibrium flux method (EFM)~\cite{efm} and the KFVS method~\cite{kfvs} can be viewed as FVS schemes in the gas kinetic version, whose intrinsic free transport mechanism makes them more suitable for solving shock waves, with the positivity preserving property~\cite{tao} and fulfilling the entropy condition~\cite{phdxu}. However, like other FVS schemes, they are not good at solving boundary layers. The TTT method mentioned in Ref.~\cite{phdxu} constructs a Maxwellian equilibrium state at the cell interface, leading to suitable viscosity, but it is not a shock capture scheme.

As proposed in Ref.~\cite{ws}, given the mean collision time and the characteristic time of flows, the direct modeling philosophy couples the macroscopic and microscopic mechanisms together, which dominates the continuum flows and rarefied flows, respectively, and their weights are determined by these two time scales. The multiscale schemes based on this direct modeling have good validity and accuracy in simulating both rarefied and continuum flows~\cite{gks,ugks,dugks,mdugks,ugkwp,cyp,mdvm}, combined with different kinds of technologies~\cite{unstructured,ib,multigrid,yuan}, which have succeeded in extending to more physics, like turbulence~\cite{liji}, flux-structure interaction~\cite{wangyong}, plasma~\cite{plasma}, multiphase flows~\cite{zeren}, non-equilibrium multi-temperature flows~\cite{caoguiyu} and so on. Before the proposition of direct modeling, it was already implied in the numerical flux of gas-kinetic method~\cite{gks}. The recently proposed KIF method~\cite{shapre} is a kind of inviscid flux extracted from the gas-kinetic schemes, coupling the KFVS and the TTT together (KFVS dominates the discontinuous part of the flow field, and TTT dominates the smooth part), aiming at predicting continuum flows in all speed regimes. Compared with the original gas-kinetic schemes, the KIF has a less computational cost, and is rather easier to be realized (imbedded) in classical CFD frameworks. Compared with the macroscopic schemes based on the Euler and the Navier-Stokes equations, the KIF has some advantages benefitting from its multiscale mechanism, like better treatment against shock instability phenomenon. The KIF can also be categorized into a series of schemes which couples the KFVS with other central-like schemes~\cite{sunshu,ohwada2018}.

Since an implicit scheme is in strong demand in many numerical experiments about flows, as well as practical flow predictions, a series of methods are developed for the numerical methods based on the gas-kinetic theory. Approximate factorization-alternating direction implicit (AF-ADI) method was used in an implicit gas-kinetic Bhatnagar-Gross-Krook (BGK) scheme for compressible flows~\cite{chit}. Xu and Mao~\cite{xu2005} developed an implicit multidimensional gas-kinetic BGK scheme for hypersonic viscous flows, based on the Lower-Upper Symmetric-Gauss-Seidel (LU-SGS)~\cite{lusgs} method. Also based on the LU-SGS, Li W. et al. ~\cite{lwd} proposed a kind of unstructured implicit gas-kinetic BGK scheme for high temperature equilibrium gas, and Li J. et al. ~\cite{liji2} extended it for unsteady flow simulations, using the dual time-stepping strategy. In this work, the LU-SGS method is used as a preconditioner of the linear system for the present method, and the generalized minimal residual (GMRES) method~\cite{gmres} is used to solve it, leading to good computational efficiency in both subsonic and hypersonic flow predictions. Mechanisms are analyzed about what influence the CFL number in this implicit scheme, and a new weight, about how to combine the KFVS well with the TTT, is proposed in order to increase the accuracy, the robustness and the smoothness between the two schemes, through the using of Ma and $\Delta P$, and smoothness is protected by the function $F_S(x,a,b)$. A series of test cases, including the lid-driven cavity flow, the hypersonic viscous flow past a cylinder, the laminar shock boundary layer interaction, the supersonic flow around a ramp segment, the hypersonic double-cone flow and the hypersonic double-ellipsoid flow, are simulated to validate the ability of the present method for solving continuum flows in all speed regimes.

The remainder of this paper is organized as the following: the present KIF method is proposed in in \ref{sec:kif}, along with a short analysis about it. \ref{sec:lusgs} is the construction of the present implicit strategy. The numerical test cases are conducted in \ref{sec:cases}. The conclusions are in \ref{sec:conclusion}.

\section{\label{sec:kif}Kinetic Inviscid Flux}
In the gas-kinetic theory, the physical system is described by the particle distribution function $f\left({\bm{x}},{\bm{\xi}},{\bm{\eta}},t\right)$, which means the number density of particles that arrive at position $\bm{x}$ at time t with velocity $\bm{\xi}$ and internal energy represented by the equivalent velocity $\bm{\eta}$. The macroscopic conservative variables can be obtained by the following relations:
\begin{equation}\label{Eq1}
\begin{aligned}
\rho &= \left \langle mf \right \rangle,\\
\rho U &= \left \langle m\xi _1f \right \rangle,\\
\rho V &= \left \langle m\xi _2f \right \rangle,\\
\rho W &= \left \langle m\xi _3f \right \rangle,\\
\rho E &= \left \langle \frac{1}{2}m(\bm{\xi}\cdot \bm{\xi}+\bm{\eta}\cdot \bm{\eta})f \right \rangle,
\end{aligned}
\end{equation}
where $m$ is the mass of a particle, $\rho$  is density, $U$, $V$, $W$ are velocity components, $E$ is energy density, and the operator $\left\langle \cdot \right\rangle$ is defined as the integral over the three-dimensional $\bm{\xi}$ and the D-dimensional $\bm{\eta}$ in the following form:
\begin{equation}
\left\langle \cdot \right\rangle = \int_{{R^3}} d\bm{\xi}\int_{{R^D}} d\bm{\eta}.
\end{equation}
This integral is called "moment" in the gas-kinetic theory. Ref.~\cite{gks} can be referred to, for details about calculating moments. Eq.~\eqref{Eq1} can be abbreviated to:
\begin{equation}
\bm{\Psi} = \left \langle \bm{\varphi}f \right \rangle.
\end{equation}
Also, the macroscopic flux $\bm{F}$ at the cell interface in the FVM framework is obtained by:
\begin{equation}
\bm{F} = \left \langle (\bm{\xi}\cdot\bm{n})\bm{\varphi}f \right \rangle.
\end{equation}
where $\bm{n}$ is the unit normal vector of the interface. Aiming at getting inviscid flux, $f$ can be approximated by Maxwellian equilibrium distribution in the following form:
\begin{equation}
f = g = {\frac{1}{2\pi P/\rho}}^{\frac{D+3}{2}}\exp[-\frac{(\bm{U}-\bm{\xi})\cdot (\bm{U}-\bm{\xi})+\bm{\eta}\cdot \bm{\eta}}{2P/\rho}],
\end{equation}
where $\bm{U}={(U,V,W)}^T$ is the velocity vector.

\subsection{The KFVS method}
The mechanism of the KFVS is that the molecules transporting across the interface to the right are from the left cell, and vice versa. Therefore, the distribution function at the cell interface is made up of two halves of the Maxwellian equilibrium distributions where the right half distribution is from the left side (cell), and vice versa:
\begin{equation}
f = \left\{
\begin{aligned}
g_L,\qquad\bm{\xi}\cdot\bm{n}\geq 0,\\
g_R,\qquad\bm{\xi}\cdot\bm{n}< 0.\\
\end{aligned}
\right.
\end{equation}
The implied physical process in the KFVS at the cell interface is the free (collisionless) molecular transportation, which is close to the process in discontinuities such as shock waves, bringing about an extra artificial viscosity. The KFVS has the positivity preserving property and fulfills the entropy condition, but like other FVS methods, it is not good at solving boundary layers. The moment equation for getting the flux of the KFVS is:
\begin{equation}
\bm{K} = {\left \langle (\bm{\xi}\cdot\bm{n})\bm{\varphi}g_L \right \rangle}_R+{\left \langle (\bm{\xi}\cdot\bm{n})\bm{\varphi}g_R \right \rangle}_L.
\end{equation}
where the half moment ${\left\langle \cdot \right\rangle}_L$ and ${\left\langle \cdot \right\rangle}_R$ are defined as:
\begin{equation}
\begin{aligned}
{\left\langle \cdot \right\rangle}_L &= \int_{-\infty}^0 d{\xi '}_1 \int_{-\infty}^{+\infty} d{\xi '}_2 \int_{-\infty}^{+\infty} d{\xi '}_3 \int_{R^D} d\bm{\eta},\\
{\left\langle \cdot \right\rangle}_R &= \int_0^{+\infty} d{\xi '}_1 \int_{-\infty}^{+\infty} d{\xi '}_2 \int_{-\infty}^{+\infty} d{\xi '}_3 \int_{R^D} d\bm{\eta}.
\end{aligned}
\end{equation}
where ${\xi '}_1=\bm{\xi}\cdot\bm{n}$, and $\bm{\xi '}$ is $\bm{\xi}$ rotated to the direction of $\bm{n}$.  Ref.~\cite{sunshu} can be referred to, for details about the rotation. The explicit KFVS is as follows:
\begin{equation}
\begin{aligned}
{K_{1}} =& \frac{{\rm{1}}}{{\rm{2}}}\left( {{\rho _L}{U_L} + {\rho _R}{U_R}} \right) + \frac{1}{2}\left( {{\rho _L}{U_L}{\chi _L} - {\rho _R}{U_R}{\chi _R}} \right) + \frac{1}{2}\left( {{\rho _L}{\theta _L} - {\rho _R}{\theta _R}} \right),\\
{K_{2}} =& \frac{{\rm{1}}}{{\rm{2}}}\left[ {\left( {{\rho _L}U_L^2 + {P_L}} \right) + \left( {{\rho _R}U_R^2 + {P_R}} \right)} \right] \\
&+ \frac{{\rm{1}}}{{\rm{2}}}\left[ {\left( {{\rho _L}U_L^2 + {P_L}} \right){\chi _L} - \left( {{\rho _R}U_R^2 + {P_R}} \right){\chi _R}} \right] + \frac{{\rm{1}}}{{\rm{2}}}\left( {{\rho _L}{U_L}{\theta _L} - {\rho _R}{U_R}{\theta _R}} \right),\\
{K_{3}} =& \frac{{\rm{1}}}{{\rm{2}}}\left( {{\rho _L}{U_L}{V_L} + {\rho _R}{U_R}{V_R}} \right) + \frac{1}{2}\left( {{\rho _L}{U_L}{V_L}{\chi _L} - {\rho _R}{U_R}{V_R}{\chi _R}} \right) + \frac{1}{2}\left( {{\rho _L}{V_L}{\theta _L} - {\rho _R}{V_R}{\theta _R}} \right),\\
{K_{4}} =& \frac{{\rm{1}}}{{\rm{2}}}\left( {{\rho _L}{U_L}{W_L} + {\rho _R}{U_R}{W_R}} \right) + \frac{1}{2}\left( {{\rho _L}{U_L}{W_L}{\chi _L} - {\rho _R}{U_R}{W_R}{\chi _R}} \right) + \frac{1}{2}\left( {{\rho _L}{W_L}{\theta _L} - {\rho _R}{W_R}{\theta _R}} \right),\\
{K_{5}} =& \frac{{\rm{1}}}{{\rm{2}}}\left( {{\rho _L}{U_L}{H_L} + {\rho _R}{U_R}{H_R}} \right)\\
&+ \frac{1}{2}\left( {{\rho _L}{U_L}{H_L}{\chi _L} - {\rho _R}{U_R}{H_R}{\chi _R}} \right) + \frac{1}{2}\left[ {\left( {{\rho _L}{H_L} - \frac{{{P_L}}}{2}} \right){\theta _L} - \left( {{\rho _R}{H_R} - \frac{{{P_R}}}{2}} \right){\theta _R}} \right],
\end{aligned}
\end{equation}
where $H=E+P/\rho$ ($P$ is the pressure) is the enthalpy density, and the subscripts "$L$" and "$R$" represent the two sides of the cell interface. $\chi$ and $\theta$ are defined as:
\begin{equation}
\begin{aligned}
{\chi}_{L/R} &= \rm{erf}(\frac{U_{L/R}}{\sqrt{2P_{L/R}/{\rho}_{L/R}}}),\\
{\theta}_{L/R} &= \sqrt{\frac{2P_{L/R}/{\rho}_{L/R}}{\pi}}\exp(\frac{-U_{L/R}^2}{\sqrt{2P_{L/R}/{\rho}_{L/R}}}).
\end{aligned}
\end{equation}
where the erf() function is the error function, and the $\exp()$ function is the exponent function.

\subsection{The TTT method}
The TTT can solve boundary layers accurately, by setting the distribution at the cell interface an exact Maxwellian one, which is obtained by making the two halves in the KFVS collide sufficiently as the following equations:
\begin{equation}\label{Eq2}
\bm{\Psi}_0 = {\left \langle \bm{\varphi}g_L \right \rangle}_R+{\left \langle \bm{\varphi}g_R \right \rangle}_L,
\end{equation}
\begin{equation}
f_0 = g(\rho_0,\bm{U}_0,P_0).
\end{equation}
Eq.~\eqref{Eq2} is used for obtaining the conservative macroscopic variables corresponding to the two half distributions. When using these conservative macroscopic variables to determine a Maxwellian distribution, it implies that the two distribution is merged into an equilibrium one. The numerical flux of TTT can be calculated from:
\begin{equation}
\bm{G} = {\left \langle (\bm{\xi}\cdot\bm{n})\bm{\varphi}f_0 \right \rangle}.
\end{equation}
Its explicit macroscopic expression is:
\begin{equation}
\begin{aligned}
\overline \rho   &= \frac{1}{2}\left( {{\rho _L}+{\rho _R}} \right) + \frac{1}{2}\left( {{\rho _L}{\chi _L} - {\rho _R}{\chi _R}} \right),\\
\overline {\rho u}  &= \frac{1}{2}\left( {{\rho _L}{U_L}+{\rho _R}{U_R}} \right) + \frac{1}{2}\left( {{\rho _L}{\theta _L} - {\rho _R}{\theta _R}} \right) + \frac{1}{2}\left( {{\rho _L}{U_L}{\chi _L} - {\rho _R}{U_R}{\chi _R}} \right),\\
\overline {\rho v}  &= \frac{1}{2}\left( {{\rho _L}{V_L}+{\rho _R}{V_R}} \right) + \frac{1}{2}\left( {{\rho _L}{V_L}{\chi _L} - {\rho _R}{V_R}{\chi _R}} \right),\\
\overline {\rho w}  &= \frac{1}{2}\left( {{\rho _L}{W_L}+{\rho _R}{W_R}} \right) + \frac{1}{2}\left( {{\rho _L}{W_L}{\chi _L} - {\rho _R}{W_R}{\chi _R}} \right),\\
\overline {\rho E}  &= \frac{1}{2}\left( {{\rho _L}{E_L}+{\rho _R}{E_R}} \right) + \frac{1}{4}\left( {{\rho _L}{U_L}{\theta _L} - {\rho _R}{U_R}{\theta _R}} \right) + \frac{1}{2}\left( {{\rho _L}{E_L}{\chi _L} - {\rho _R}{E_R}{\chi _R}} \right).
\end{aligned}
\end{equation}
Then the TTT flux is the simple Euler flux, using the merged macroscopic variables, as follows:
\begin{equation}
\begin{aligned}
{G_{1}} &= \bar \rho \bar U,\\
{G_{2}} &= \bar \rho {{\bar U}^2} + \bar P,\\
{G_{3}} &= \bar \rho \bar U\bar V,\\
{G_{4}} &= \bar \rho \bar U\bar W,\\
{G_{5}} &= \bar \rho \bar U\bar H,
\end{aligned}
\end{equation}
where $\bar U=\overline{\rho U}/\bar \rho$, $\bar V=\overline{\rho V}/\bar \rho$, $\bar W=\overline{\rho W}/\bar \rho$ ,$\bar e =\overline{\rho E}/\bar \rho -  \frac{1}{2}({\bar U}^2+{\bar V}^2+{\bar W}^2)$, $\bar P =(\gamma-1)\bar \rho \bar e$, $\bar H =\overline{\rho E}/\bar \rho +\bar P /\bar \rho$. $\gamma$ is the ratio of specific heat, which is often set 1.4 for air.

The TTT is an accurate scheme for boundary layers simulations. While, it is not a shock capture scheme, since the distribution function in the shock wave is in non-equilibrium state from the microscopic point of view, and from the macroscopic point of view, there is no numerical viscosity at all associated with the equilibrium Maxwellian distribution for capturing shock waves.

\subsection{The present KIF method}
In the KIF method, the numerical flux is written as:
\begin{equation}
\bm{F} = \beta \bm{K}+(1-\beta)\bm{G},
\end{equation}
where $\beta$ is a kind of weight (the weight of free transport mechanism at a cell interface). Among studies about $\beta$, the high fidelity physical mechanism~\cite{ohwada2006,ohwada2018,shapre} and a simple mathematical form~\cite{sunshu} are taken into consideration. In this study, the smoothness expression of $\beta$ is considered (the first derivative should be continuous), which brings about a slightly complex formula, but better convergence property. The new expression for $\beta$ is:
\begin{equation}\label{Eq3}
\beta = F_S(\Delta P,0,1)F_S(Ma,C_2 Ma_{back},Ma_{back}),
\end{equation}
where $F_S(x,a,b)$ function is a smooth function consisting of $\sin()$ function as follows:
\begin{equation}
F_S(x,a,b) = \left\{
\begin{aligned}
&0,\qquad  &x<a,\\
&1,\qquad  &x>b,\\
&0.5\sin(\pi\frac{x-a}{b-a}-\frac{\pi}{2})+0.5,\qquad &a\leq x\leq b.
\end{aligned}
\right.
\end{equation}
The first part of Eq.~\eqref{Eq3}, $F_S(\Delta P,0,1)$, works as a detector for discontinuity, where
\begin{equation}\label{Eq4}
\Delta P = C_1\mathop{max}\limits_{\omega\in\Omega}(\rvert\frac{P_L-P_R}{P_L+P_R}\rvert).
\end{equation}
Here, $C_1$ is set 100 as a constant. A large value of $C_1$ is chosen mainly to capture the head and the tail of a discontinuity, and give them enough $\beta$. $\Omega$ is a gather of all the surfaces of the two adjacent cells of a cell interface whose flux is to be calculated. The reason for utilizing the max() function in Eq.~\eqref{Eq4} is that, when dealing with strong shocks, the artificial viscous flux in the interface, parallel to the flow direction, should be enough in order to avoid the shock instability phenomenon. Refs.~\cite{sunshu,ohwada2018,shapre} use the similar technologies, while a smaller $C_1$ is often chosen to reduce $\beta$ in the boundary layer calculations. The $C_1$ here can use large value because of the use of $Ma$ number in the second part of Eq.~\eqref{Eq3}, which is $F_S (Ma,C_2 Ma_{back},Ma_{back})$. This term aims to reduce the $\beta$ to zero around boundary layers, and promise enough KFVS around all of the shocks. At boundaries, a large $\beta$ not only destroys accuracy but brings about serious instability and restricts the CFL number in implicit schemes as well. For example, when simulating the hypersonic viscous flow past a cylinder by the implicit KIF, the CFL number can be 600, but without these technologies, it can only be about 20 when the simulation starts. The $Ma_{back}$ is the $Ma$ number behind the normal shock calculated from the inlet $Ma$ number, $Ma_{inlet}$ , as follows:
\begin{equation}
Ma_{back} = \left\{
\begin{aligned}
&1,\qquad & Ma_{inlet}<1,\\
&\sqrt{\frac{1+[(\gamma-1)/2]Ma_{inlet}^2}{\gamma Ma_{inlet}^2-(\gamma-1)/2}},\qquad & Ma_{inlet} \geq 1.
\end{aligned}
\right.
\end{equation}
The parameter $C_2$ in $F_S(Ma,C_2 Ma_{back},Ma_{back})$ is set 0.5 and the $Ma$ number is the local $Ma$ number at the cell interface:
\begin{equation}
Ma=\max(Ma_L,Ma_R)
\end{equation}

\section{\label{sec:lusgs}Implicit Strategy}
In implicit schemes, with regard to a cell $i$, the residual $\bm{R}^{n+1}$, including the convection flux and the viscous flux, is linearized into:
\begin{equation}
\bm{R}^{n+1}=\bm{R}^n+(\frac{\partial\bm{R}}{\partial\bm{\Psi}})\Delta\bm{\Psi}^n,
\end{equation}
and the marching equation is:
\begin{equation}\label{Eq5}
[\frac{\Omega}{\Delta t}+\frac{\partial\bm{R}}{\partial\bm{\Psi}}]\Delta\bm{\Psi}^n=-\bm{R}^n,
\end{equation}
where the $\Omega$ represents the cell volume, and $\Delta t$ is the iteration time step. In this work, $\bm{R}^n$ in Eq.~\eqref{Eq5} is given by the KIF convection flux and the viscous flux, which ensure the result, and the left hand side of Eq.~\eqref{Eq5} can be obtained from other methods (the ordinary used first order Roe averaged Jacobian matrix is chosen for the convection flux in this paper), which only influences the process of convergence. When the LU-SGS method is used for solving the linear system on the unstructured mesh, Eq.~\eqref{Eq5} is solved as:
\begin{equation}
(\bm{D}+\bm{L})\bm{D}^{-1}(\bm{D}+\bm{U})\Delta\bm{\Psi}^n=-\bm{R}^n,
\end{equation}
where
\begin{equation}
\left\{
\begin{aligned}
&\bm{L}=\sum_{j \in \bm{L}(i)}\left[-\bm{A}_{\text {Roe}, j}^{-}-\bm{A}_{v, j}^{-}\right] S_{i j}, \\
&\bm{U}=\sum_{j \in \bm{U}(i)}\left[\bm{A}_{\text {Roe}, j}^{+}+\bm{A}_{v, j}^{+}\right] S_{i j}, \\
&\bm{D}=\frac{\Omega_{i}}{\Delta t_{i}} \bar{I}+\sum_{j=1}^{N_{(i)}}\left[\bm{A}_{\text {Roe}, j}^{+}+\bm{A}_{v, j}^{+}\right]S_{i j}.
\end{aligned}
\right.
\end{equation}
Here, $\bm{L}(i)$ and $\bm{U}(i)$ denote the nearest neighbors of the cell $i$ which belong to the lower and the upper matrix. $S_{ij}$ represents the area of the face shared by cell $i$ and cell $j$. $N_{(i)}$ stands for the number of faces of the control volume $\Omega_i$. $\bar{I}$ is the unit matrix. The Jacobian matrixes $\bm{A}_{\text{Roe},j}^+$, $\bm{A}_{\text{Roe},j}^-$ are from $\partial \bm{F}_{c,ij} / \partial \bm{\Psi}_i$ and $\partial \bm{F}_{c,ij} / \partial \bm{\Psi}_j$, respectively, where $\bm{F}_{c,ij}$ is the Roe convection flux from cell $i$ to cell $j$. The Jacobian matrixes $\bm{A}_{v, j}^{+}$, $\bm{A}_{v, j}^{-}$ are $\partial \bm{F}_{v,ij} / \partial \bm{\Psi}_i$ and $\partial \bm{F}_{v,ij} / \partial \bm{\Psi}_j$, respectively, where $\bm{F}_{v,ij}$ is the viscous flux from cell $i$ to cell $j$, obtained by the usual central scheme in CFD frameworks~\cite{jiri}.

When using the LU-SGS, Eq.~\eqref{Eq5} is transformed into the following two-step inversion procedure:
\begin{equation}
\left\{
\begin{aligned}
&(\bm{D}+\bm{L})\Delta\bm{\Psi}^\ast=-\bm{R}^n, \\
&(\bm{D}+\bm{U})\Delta\bm{\Psi}^n=\bm{D}\Delta\bm{\Psi}^\ast.
\end{aligned}
\right.
\end{equation}

The LU-SGS method is used as a preconditioner of the linear system ~\eqref{Eq5}, to be solved by the GMRES method. The GMRES method is an advanced iterative method for solving linear systems, which has the property of minimizing at every step the norm of the residual vector over a Krylov subspace~\cite{gmres}. The efficiency of Krylov-subspace methods depends strongly on good preconditioners, whose purpose is to cluster the eigenvalues of the system matrix around unity~\cite{jiri}. When using the LU-SGS method as a preconditioner, a significant increase in efficiency is brought out and memory is reduced, compared with other preconditioners~\cite{lg}.

The iteration time and the CFL number refers to both the viscid part and the inviscid part~\cite{su1},
\begin{equation}
\Delta t_i=N_{CFL}\min(\frac{|\Omega_i|}{\lambda_i^c},0.25\frac{{|\Omega_i|}^2}{\lambda_i^v}),
\end{equation}
\begin{equation}
\begin{aligned}
\lambda_i^c&=\sum_{j\in N_{(i)}}(|\bm{u}_{ij}\cdot\bm{n}_{ij}|+c_{ij})S_{ij},\\
\lambda_i^v&=\sum_{j\in N_{(i)}}\frac{\mu_{ij}}{\rho_{ij}}S_{ij}^2,
\end{aligned}
\end{equation}
where $N_{CFL}$ is the CFL number, $\bm{u}_{ij}=(\bm{u}_i+\bm{u}_j)/2$, and $c_{ij}=(c_i+c_j )/2$ denote the velocity vector and the speed of sound at the cell interface. $\rho_{ij}=(\rho_i+\rho_j )/2$ is the density, and,
\begin{equation}
\mu_{ij}=\frac{4}{3}(\mu_{l,ij}+\mu_{e,ij})+(1+\frac{Pr_l}{Pr_t}\frac{\mu_{l,ij}}{\mu_{e,ij}})(\gamma\frac{\mu_{l,ij}}{Pr_l}),
\end{equation}
where $\mu_{l,ij}=(\mu_{l,i}+\mu_{l,j})/2$ and $\mu_{e,ij}=(\mu_{e,i}+\mu_{e,j})/2$ are laminar and eddy viscosities. $Pr_l$ and $Pr_t$  are laminar and eddy Prandtl numbers.

In this study, the 2-norm of density between adjacent steps is chosen as residual to judge convergence.

\section{\label{sec:cases}Numerical Test Cases}
Six numerical tests are used to examine the present method. The lid-driven cavity flow and the hypersonic viscous flow past a cylinder are simulated to examine the implicit efficiency in subsonic and hypersonic flow respectively. The hypersonic viscous flow past a cylinder, the laminar shock boundary layer interaction and the supersonic flow around a ramp segment are three typical supersonic/hypersonic flows, and they are used to validate the present KIF method. The hypersonic double-cone flow and the hypersonic double-ellipsoid flow are two complex flows for validating the ability of both the KIF method and the present implicit strategy. In all these numerical test cases, the MUSCL reconstruction with venkatakrishnan limiter~\cite{venkata} is used.

\subsection{The lid-driven cavity flow}\label{Sec:caseA}
The lid-driven cavity flow is a closed incompressible flow in a square region bounded by a top moving wall towards right and other three static walls, which is simulated in this section to examine the implicit efficiency of the present method as well as its accuracy for viscous dominated flow. In this case, the $Ma$ number of the lid is set 0.1, and the $Re$ number is set 1000 whose reference length is the wall length. A Cartesian grid with 128 nodes in each wall is used.

\begin{figure}[!htp]
\centering
\includegraphics[width=0.5\textwidth]{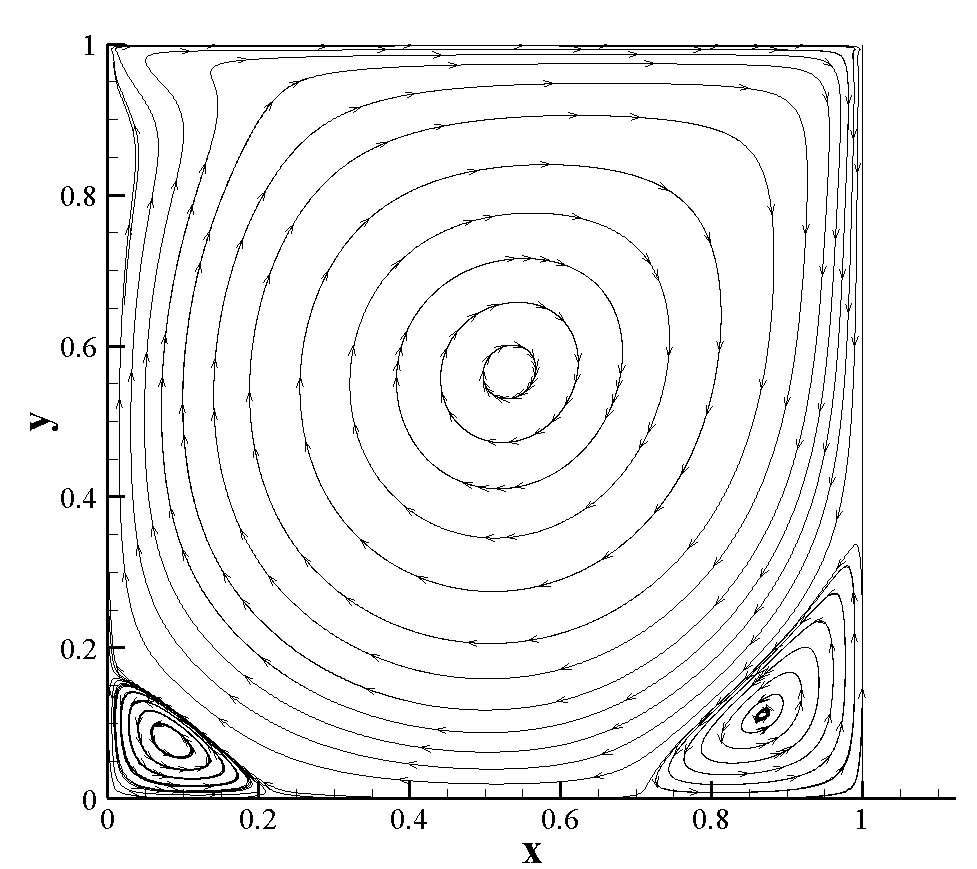}
\caption{\label{Fig:cavity1000fai} The streamline of the lid-driven flow at $Re = 1000$.}
\end{figure}

\begin{figure}[!htp]
\centering
\subfigure[]{
\includegraphics[width=0.45\textwidth]{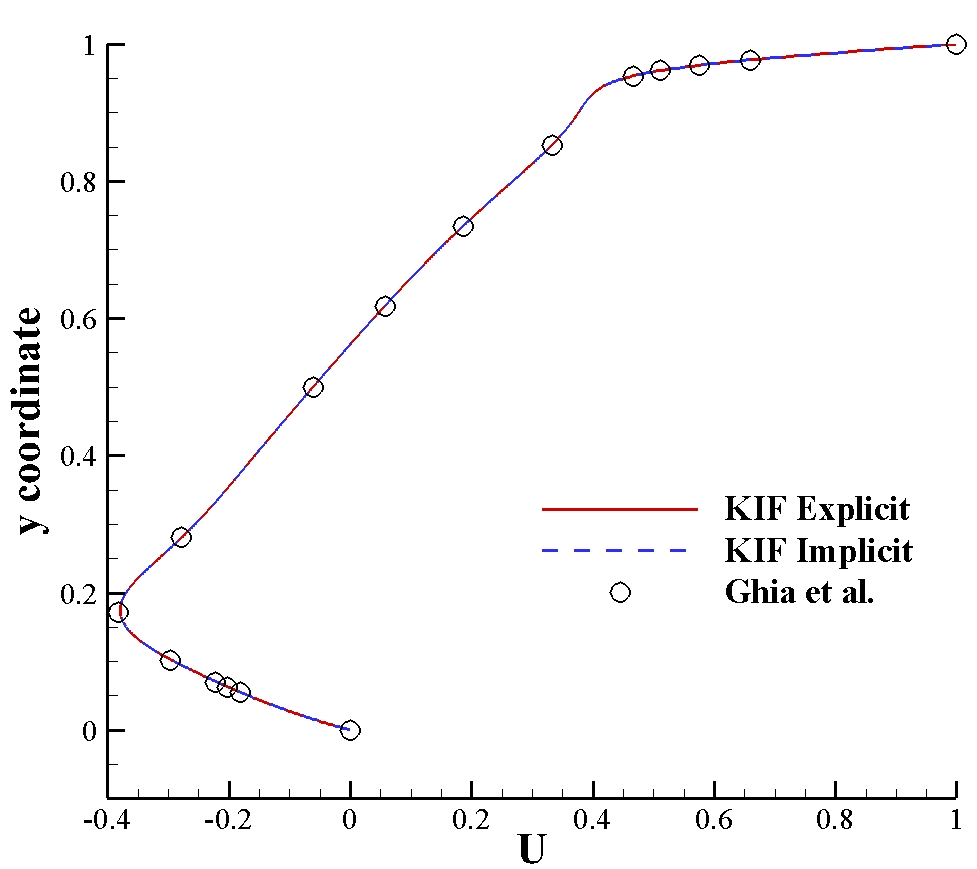}
}
\subfigure[]{
\includegraphics[width=0.45\textwidth]{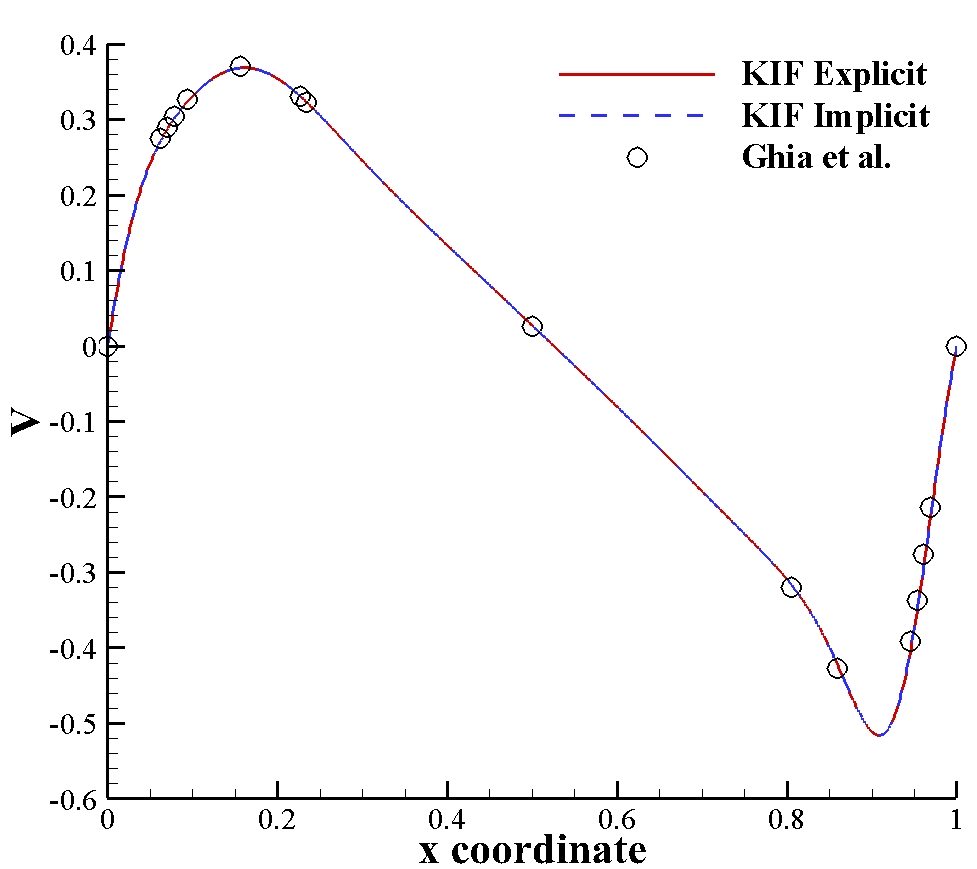}
}
\caption{\label{Fig:cavity1000kif_uv} The velocity curve: (a) The U-velocity along the central-vertical line, (b) the V-velocity along the central-horizontal line.}
\end{figure}

\begin{figure}[!htp]
\centering
\includegraphics[width=0.6\textwidth]{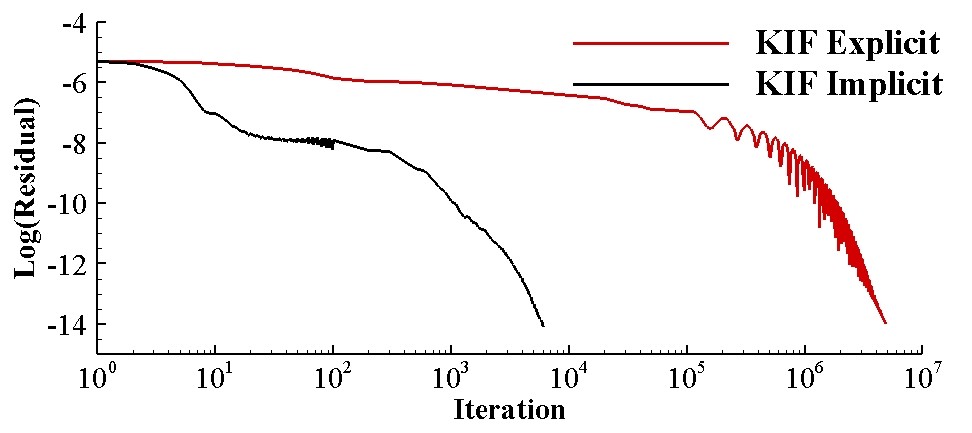}
\caption{\label{Fig:cavity1000res} The residual-time curve of the lid-driven flow at $Re = 1000$.}
\end{figure}

Both the explicit method and the implicit method are conducted for comparison, whose CFL numbers are 0.3 and 1000, respectively. When simulating subsonic flow, the CFL number of the explicit KIF method cannot be large in case of divergence. The results of Ghia et al.~\cite{ghia} are also used for comparison. The streamline calculated by the implicit KIF is shown in Fig.~\ref{Fig:cavity1000fai}. The U-velocity along the central vertical line and the V-velocity along the central horizontal line are shown in Fig.~\ref{Fig:cavity1000kif_uv}, matching well with other results. The three vortex centers in the flow field are shown in Tab.~\ref{table:vertex}. The results of the explicit KIF and the results of the implicit KIF are nearly the same. The errors between the results of the KIF and the results of Ghia et al. are less than 3\%, the maximum of which is on the bottom left-hand corner. The residual-time curve is shown in Fig.~\ref{Fig:cavity1000res}. From the beginning to convergence ($10^{-14}$), the iteration number of the explicit KIF is about 800 times of the implicit one.

\subsection{The hypersonic viscous flow past a cylinder}
In this case, a hypersonic viscous flow passes through a cylinder and forms a bowl shock in front of it. The flow can be counted as laminar one around the forehead of the cylinder. This flow is often chosen as a hypersonic benchmark test case~\cite{xu2005,pan}, to test the shock instability phenomenon like carbuncle~\cite{kk1} and to test the ability of a numerical method to predict the heat flux precisely~\cite{kk2}, especially at the stagnation point. Besides these, this case is also used to examine the implicit efficiency for the hypersonic flow in this work.

The inflow $Ma$ number is 8.03 and the $Re$ number is 183500 whose reference length is the radius of the cylinder, 1.0. The inflow temperature is 124.94K. On the cylinder, the isothermal boundary condition is used and the temperature of the wall is 294.44K. The grid is $120\times180$, and the first layer adjacent to the cylinder is set to be $1.5\times10^{-5}$. The inflow boundary of the flow domain is a half circle whose radius is 4.0.

\begin{figure}[!htp]
\centering
\subfigure[\label{Fig:bbkif_rho}]{
\includegraphics[width=0.35\textwidth]{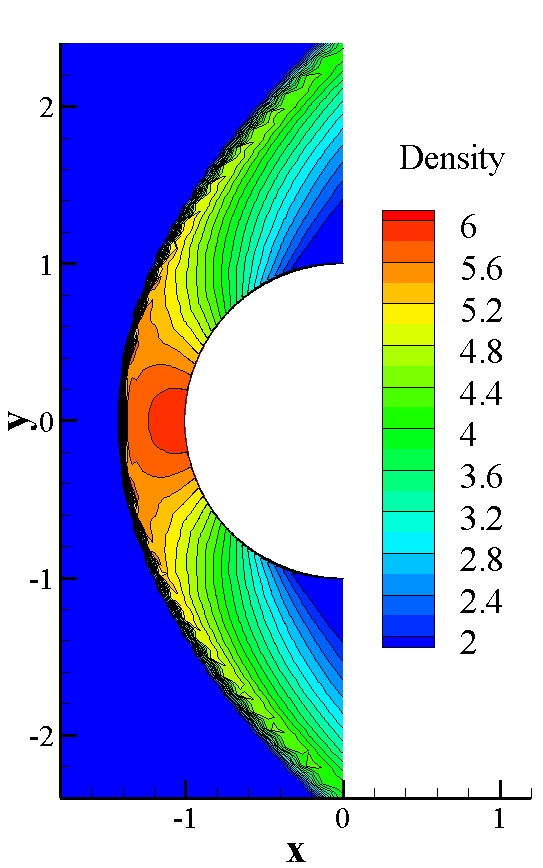}
}
\subfigure[\label{Fig:bbkif_ratiof}]{
\includegraphics[width=0.35\textwidth]{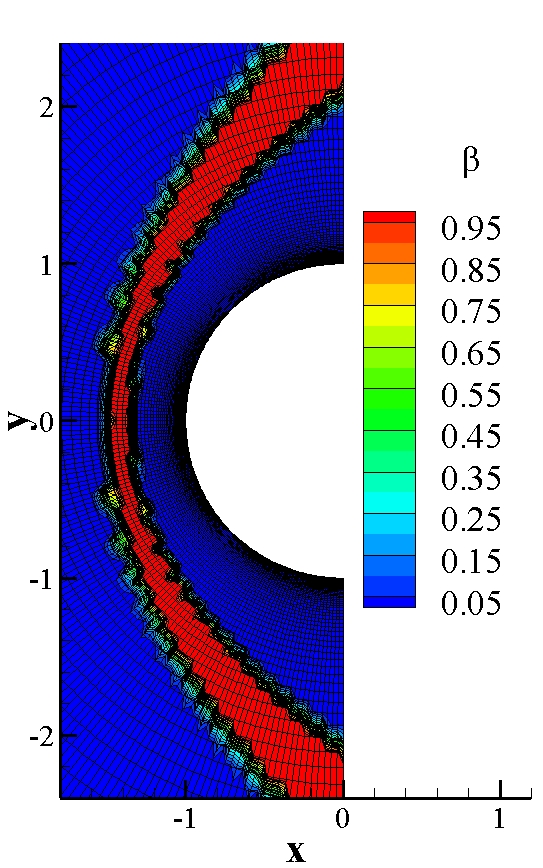}
}
\caption{The contours of (a) density and (b) $\beta$ in the hypersonic viscous flow past a cylinder.}
\end{figure}

\begin{figure}[!htp]
\centering
\subfigure[]{
\includegraphics[width=0.45\textwidth]{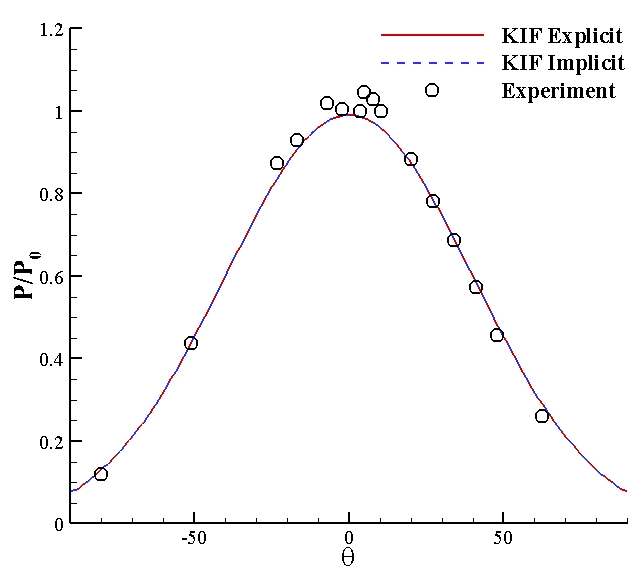}
}
\subfigure[]{
\includegraphics[width=0.45\textwidth]{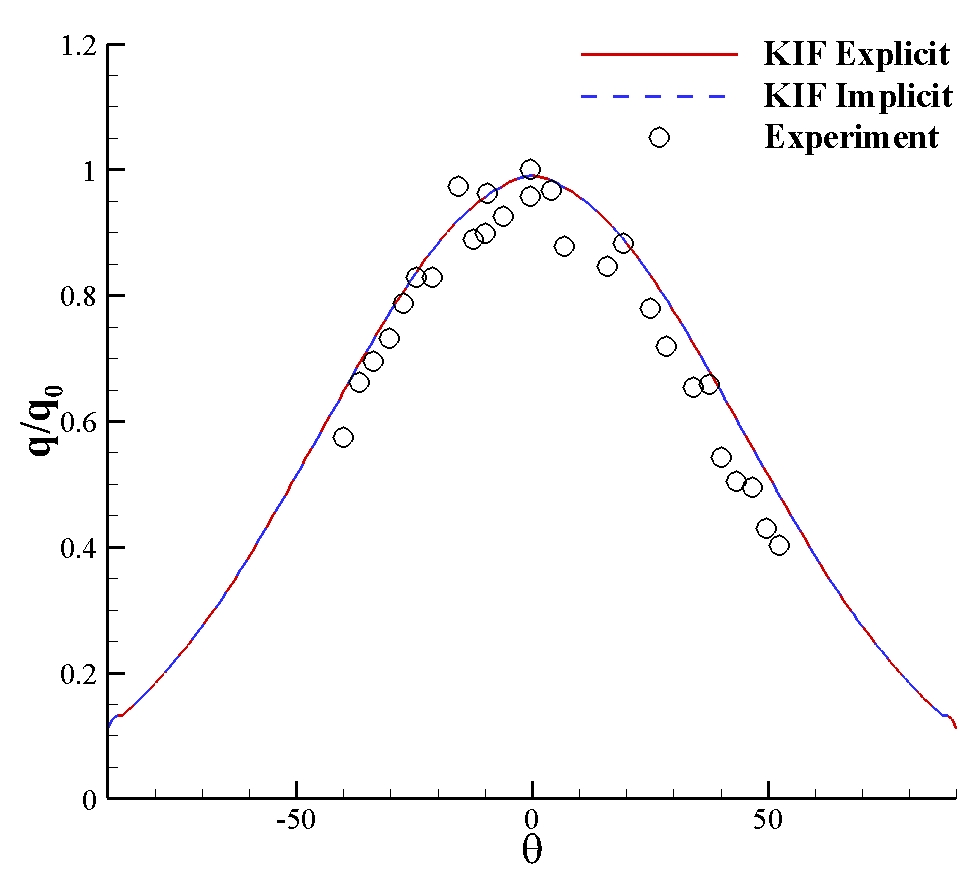}
}
\caption{\label{Fig:bbkif_pq} The curves of (a) pressure and (b) heat flux on the wall in the hypersonic viscous flow past a cylinder.}
\end{figure}

\begin{figure}[!htp]
\centering
\includegraphics[width=0.6\textwidth]{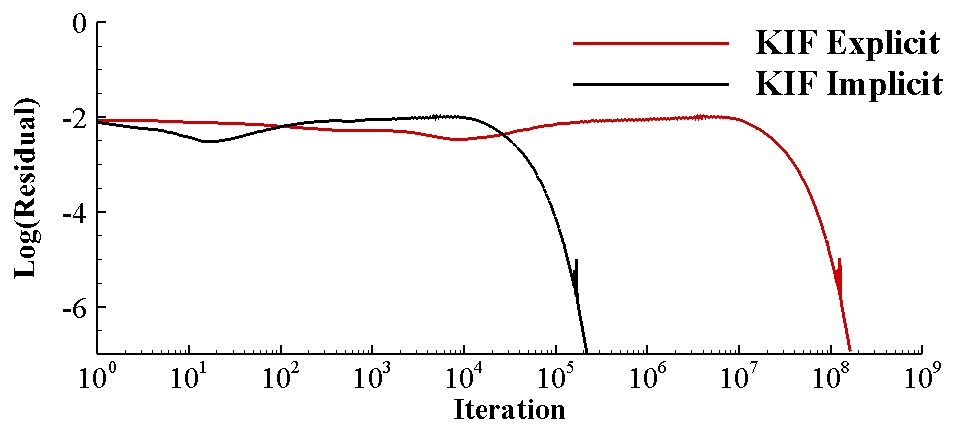}
\caption{\label{Fig:bbres} The residual-time curve of the hypersonic viscous flow past a cylinder.}
\end{figure}

Both the explicit method and the implicit method are used for comparison, whose CFL numbers are 0.8 and 600, respectively. If the KFVS part in the boundary layer is not controlled carefully, the CFL number of the implicit KIF can only be about 20 when the simulation starts. The density contour predicted by the present method is shown in Fig.~\ref{Fig:bbkif_rho}, which is clean and without the shock instability phenomenon. The $\beta$ contour predicted by the present method is shown in Fig.~\ref{Fig:bbkif_ratiof}, about how free transport mechanism is introduced in the discontinuous regime of this hypersonic flow. The pressure and heat flux curves at the surface are shown in Fig.~\ref{Fig:bbkif_pq}, normalized by $0.9209\rho_\infty U_\infty^2$ and $0.003655\rho_\infty U_\infty^3$, respectively, with the experiment data~\cite{expcylinder} for comparison. The results of the KIF match well with the experiment data. The residual-time curve is shown in Fig.~\ref{Fig:bbres}. From the beginning to convergence (residual is $1.0\times10^{-6.3}$), the iteration number of the explicit KIF is about 850 times of the implicit one, which is very similar to the result of Sec.~\ref{Sec:caseA}.

\subsection{The laminar shock boundary layer interaction}
This test case deals with the interaction of an oblique shock with a laminar boundary layer on an adiabatic wall, which is a benchmark~\cite{ohwada2006}, examining the numerical method for solving such phenomenon, the separated flow in the boundary layer caused by the shock wave. The flow domain is $(-0.1\leq x\leq 1.6)\times(0\leq y\leq 1)$. The angle of the shock is $32.6^{\circ}$, hitting the wall at $x = 1$, and the wall begins at $x = 0$. The $Ma$ number of the shock is 2 and the $Re$ number is $2.96 \times 10^5$ whose reference length is 1.0. The mesh is $120\times 100$. The minimum of the cell height is $1 \times 10^{-5}$.

\begin{figure}[!htp]
\centering
\includegraphics[width=0.6\textwidth]{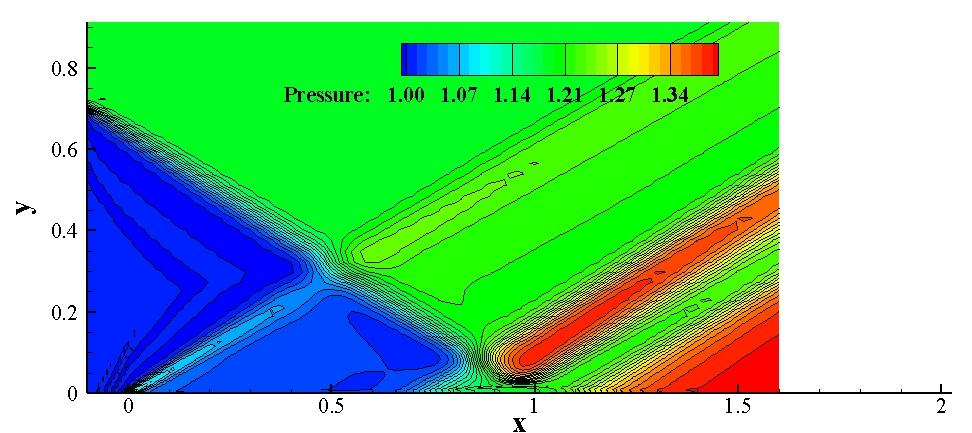}
\caption{\label{Fig:lblkif_p} The pressure contour of the laminar shock boundary layer interaction.}
\end{figure}

\begin{figure}[!htp]
\centering
\subfigure[\label{Fig:lblkif_line_p}]{
\includegraphics[width=0.45\textwidth]{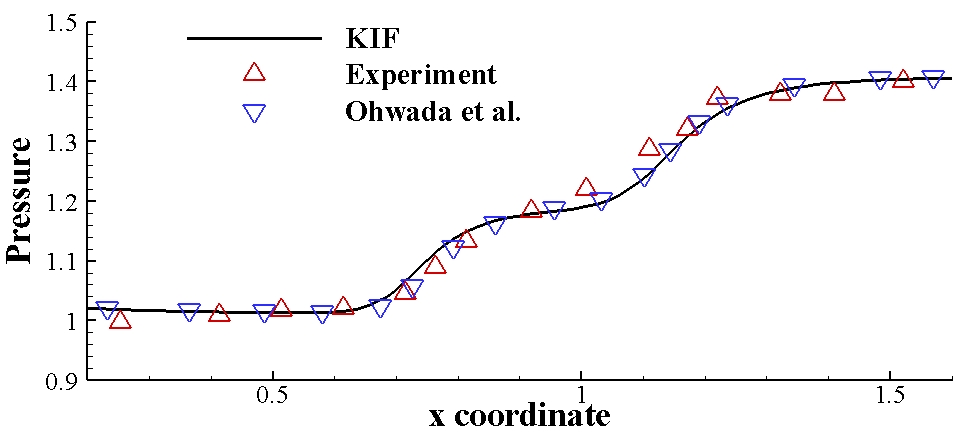}
}
\subfigure[\label{Fig:lblkif_line_cf}]{
\includegraphics[width=0.45\textwidth]{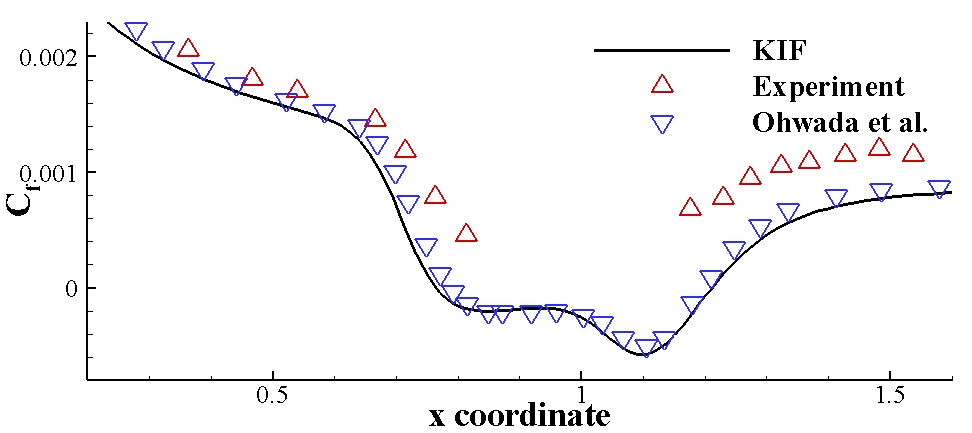}
}
\caption{\label{Fig:lblkif_line_pcf} The curves of (a) pressure and (b) friction coefficient on the wall in the laminar shock boundary layer interaction.}
\end{figure}

When simulating this flow, the CFL number is set 2000 and the iteration number is $3 \times 10^5$. The pressure contour is shown in Fig.~\ref{Fig:lblkif_p}. The pressure and friction coefficient along the wall are shown in Fig.~\ref{Fig:lblkif_line_pcf}, which agree well with the experiment data~\cite{explbl} and the results of Ref.~\cite{ohwada2006}. In Fig.~\ref{Fig:lblkif_line_p}, the resultant pressure platform, caused by the locally separated flow, is clear.

\subsection{The supersonic flow around a ramp segment}
In this test case, the supersonic flow around a ramp segment with inclined angles $10^{\circ}$ and $-25^{\circ}$ is considered. When the angle is $10^{\circ}$, it is a compression (ramp) flow, and when the angle is $-25^{\circ}$, it is an expansion (corner) flow. Both of them are typical supersonic flows~\cite{shu2016,2000Parabolized}. The $Ma$ number is 3 and the $Re$ number is 16800 whose reference length is 1.0. The inflow temperature is 216.7K. On the ramp segment, the isothermal boundary condition is used and the temperature of the wall is 606.7K. The mesh is $220\times 60$. The minimum of the cell height is $4 \times 10^{-4}$.

\begin{figure}[!htp]
\centering
\subfigure[\label{Fig:maramp10}]{
\includegraphics[width=0.45\textwidth]{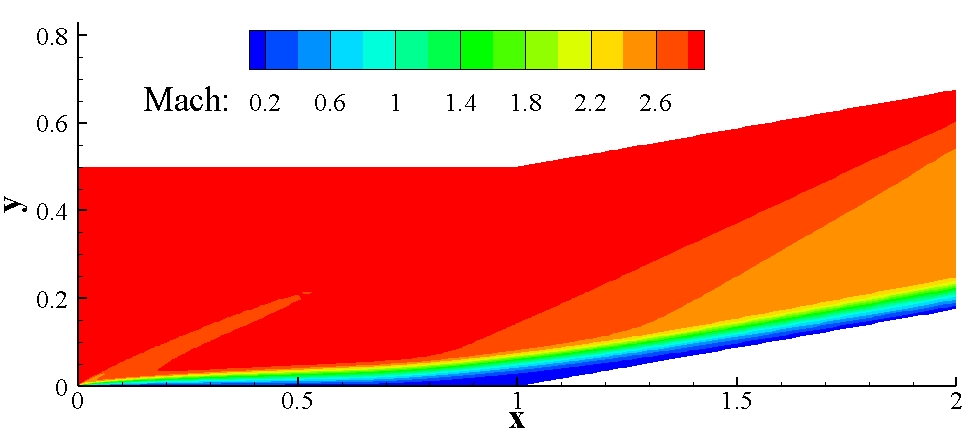}
}
\subfigure[\label{Fig:maramp25}]{
\includegraphics[width=0.45\textwidth]{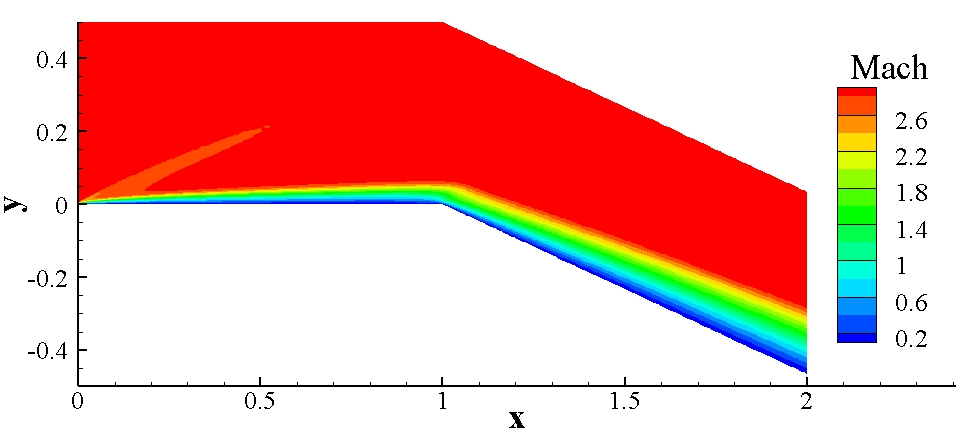}
}
\caption{\label{Fig:maramp} The $Ma$ contours of the supersonic flow around a ramp segment: (a) The $10^{\circ}$ compression ramp flow, (b) the $-25^{\circ}$ expansion corner flow.}
\end{figure}

\begin{figure}[!htp]
\centering
\subfigure[\label{Fig:pramp10}]{
\includegraphics[width=0.45\textwidth]{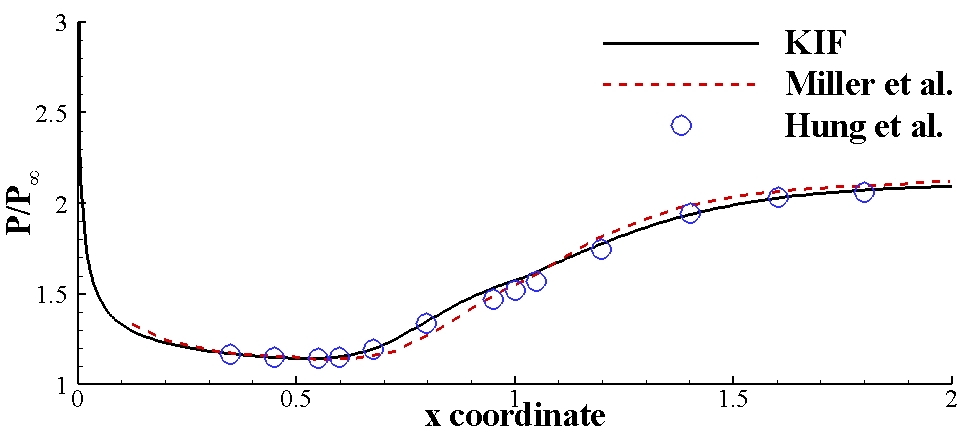}
}\hspace{0.05\textwidth}%
\subfigure[\label{Fig:cframp10}]{
\includegraphics[width=0.45\textwidth]{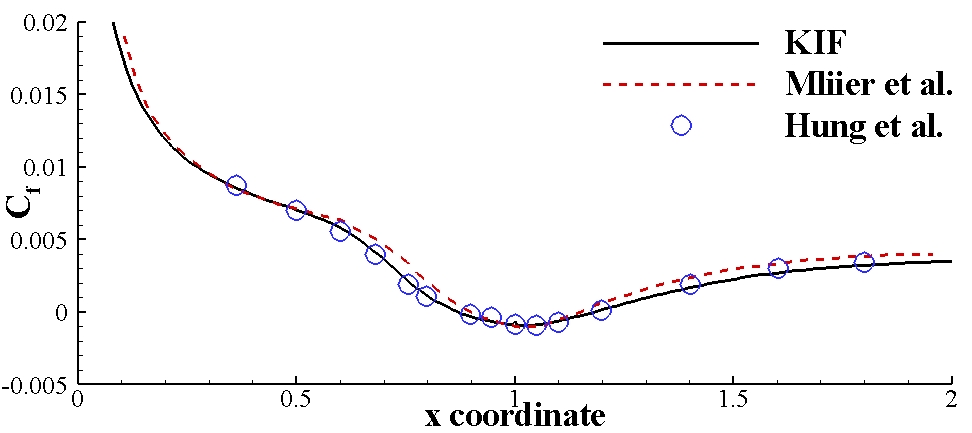}
}\\
\subfigure[\label{Fig:pramp25}]{
\includegraphics[width=0.45\textwidth]{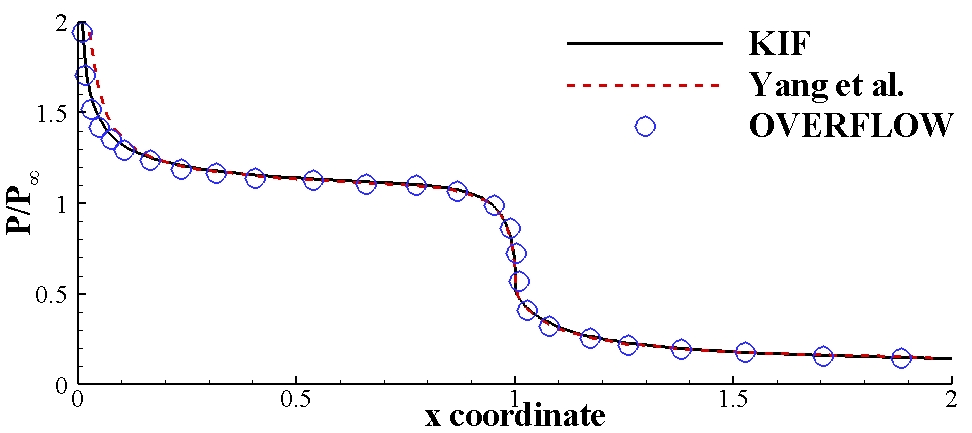}
}\hspace{0.05\textwidth}%
\subfigure[\label{Fig:cframp25}]{
\includegraphics[width=0.45\textwidth]{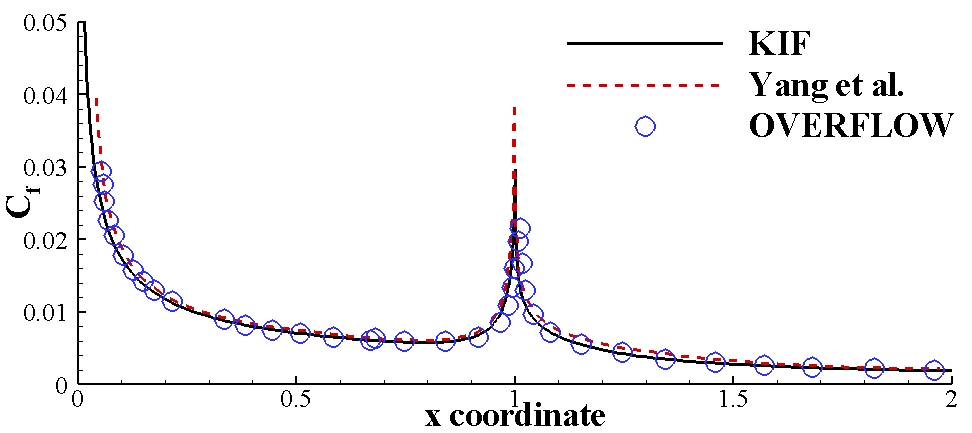}
}
\caption{\label{Fig:pcframp} The pressure and friction coefficient curves on the wall of the supersonic flow around a ramp segment: (a) The pressure curve of the $10^{\circ}$ compression ramp flow, (b) the friction coefficient curve of the $10^{\circ}$ compression ramp flow, (c) the pressure curve of the $-25^{\circ}$ expansion corner flow, (d) the friction coefficient curve of the $-25^{\circ}$ expansion corner flow.}
\end{figure}

The CFL number is set 2000 and the convergence iteration number is 5000. The $Ma$ number contour predicted by the present method is shown in Fig.~\ref{Fig:maramp}. The pressure and friction coefficient along the wall are shown in Fig.~\ref{Fig:pcframp}. The results of Miller et al.~\cite{2000Parabolized}, the results of Huang et al.~\cite{maccormark}, the results of Yang et al.~\cite{shu2016} and the results of OVERFLOW proved by Miller et al.~\cite{2000Parabolized} are used for comparison. The results predicted by the present method match well with them, validating that the present method can simulate the supersonic flows around the compression ramp and the expansion corner accurately.

\subsection{The hypersonic double-cone flow}
This test case is a complex three dimensional hypersonic test case and axisymmetric code is used. The half-angle of the first cone is $25^{\circ}$ and that of the second cone is $55^{\circ}$. A hypersonic flow passes, bringing about an attached shock wave originating at the first cone tip, a detached shock wave formed by the second cone, and a resulting shock triple point. The transmitted shock impinges on the second-cone surface, which separates the flow and produces a large localized increase in the pressure and heat-transfer rate. This pressure rise causes the flow to separate and also produces a supersonic underexpanded jet that flows downstream near the second-cone surface.~\cite{babinsky}

In this case, the inflow $Ma$ number is 9.59 and the $Re$ number is 130900 whose reference length is 1.0. The inflow temperature is 185.56K. On the double cone, the isothermal boundary condition is used and the temperature of the wall is 293.33K. Working gas is chosen as nitrogen. The element number of the mesh is 164106 and the first layer adjacent to the model is $1.0\times10^{-6}$.

\begin{figure}[!htp]
\centering
\subfigure[]{
\includegraphics[width=0.28\textwidth]{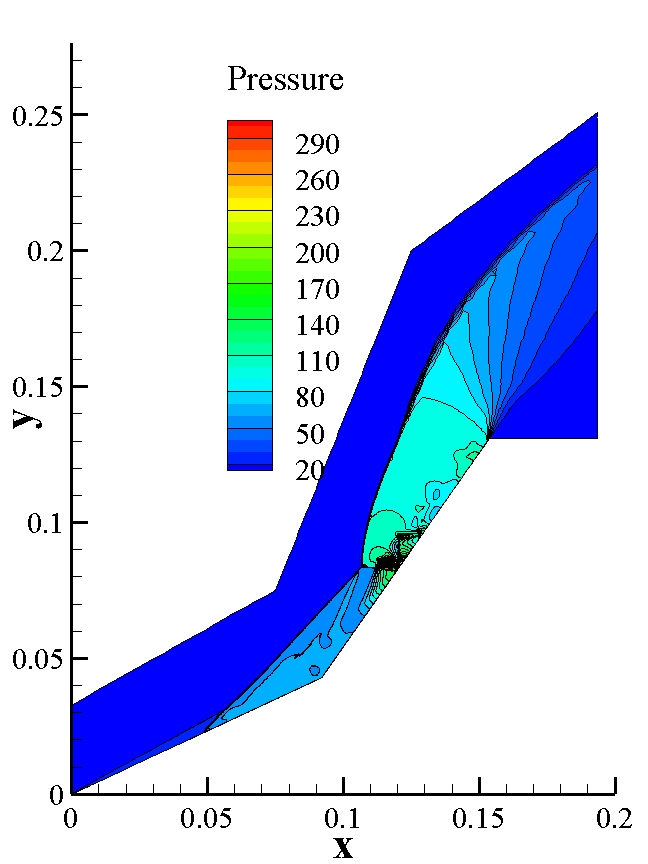}
}\hspace{0.02\textwidth}%
\subfigure[]{
\includegraphics[width=0.28\textwidth]{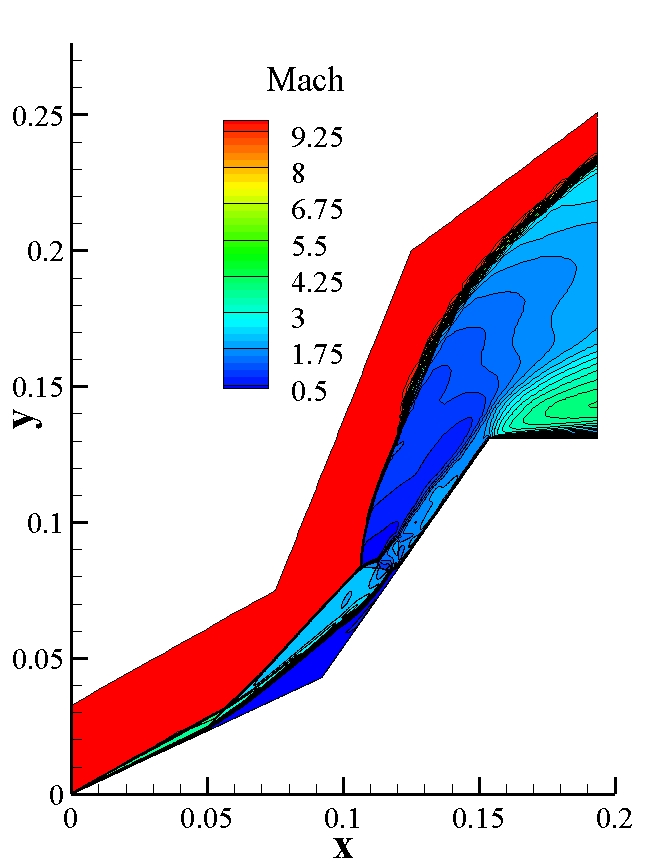}
}
\subfigure[]{
\includegraphics[width=0.28\textwidth]{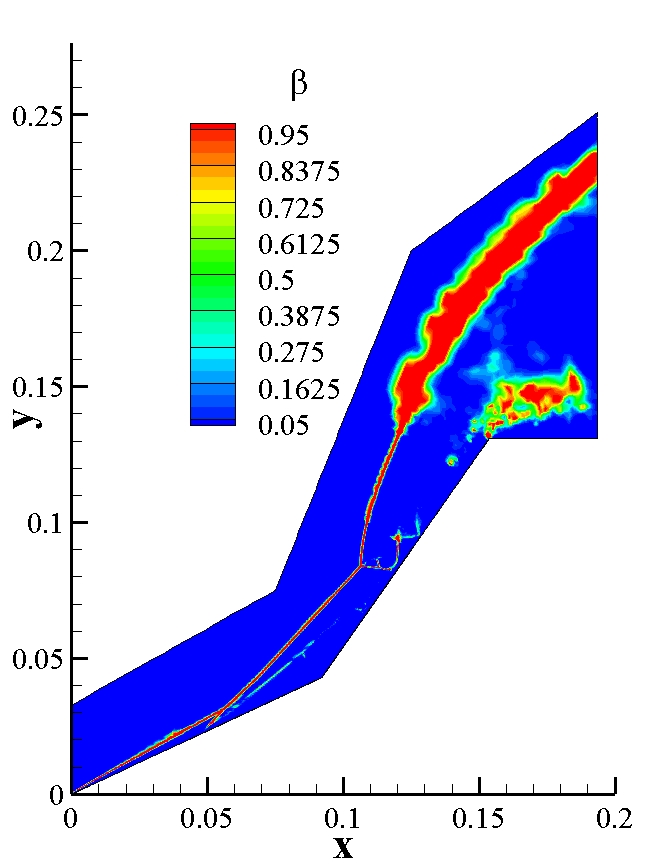}
}\\
\subfigure[]{
\includegraphics[width=0.28\textwidth]{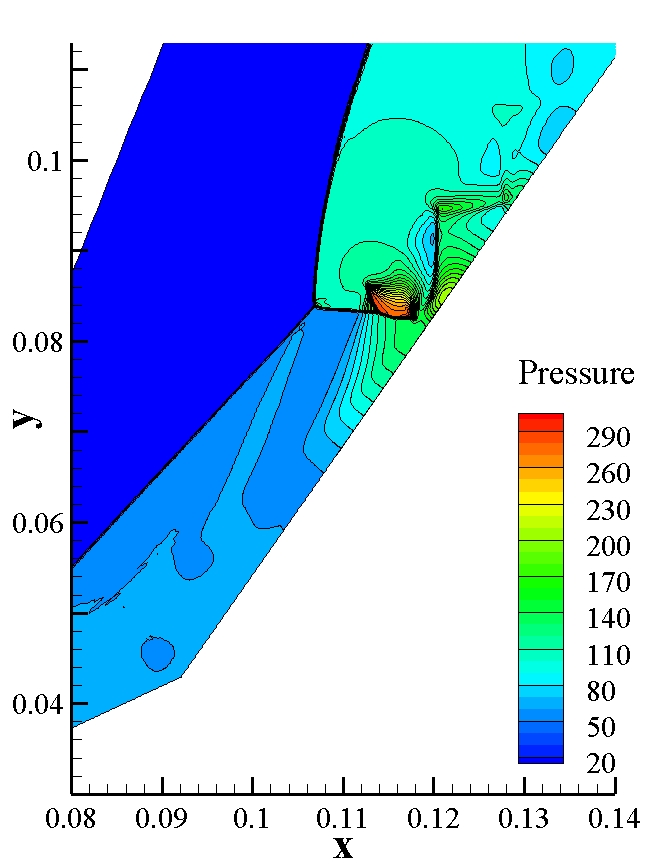}
}\hspace{0.02\textwidth}%
\subfigure[\label{Fig:hdckif_ma3}]{
\includegraphics[width=0.28\textwidth]{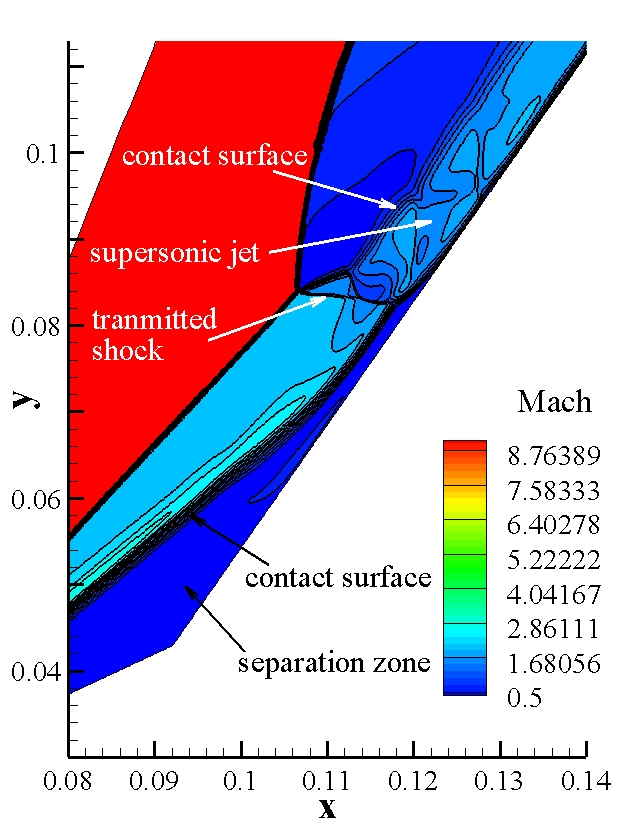}
}
\subfigure[\label{Fig:hdckif_ratiof2}]{
\includegraphics[width=0.28\textwidth]{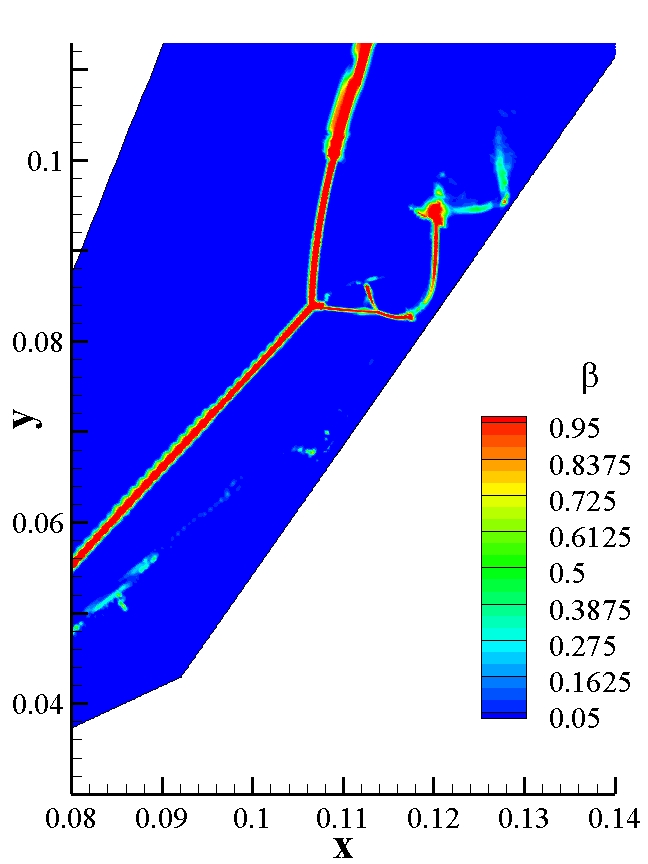}
}
\caption{\label{Fig:hdckif_map} The pressure contours, $Ma$ contours and $\beta$ contours of the hypersonic double-cone flow.}
\end{figure}

\begin{figure}[!htp]
\centering
\subfigure[]{
\includegraphics[width=0.45\textwidth]{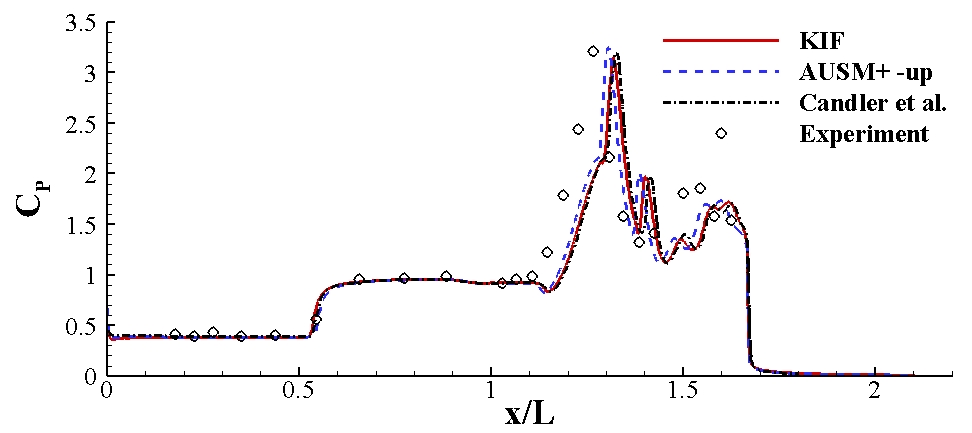}
}
\subfigure[]{
\includegraphics[width=0.45\textwidth]{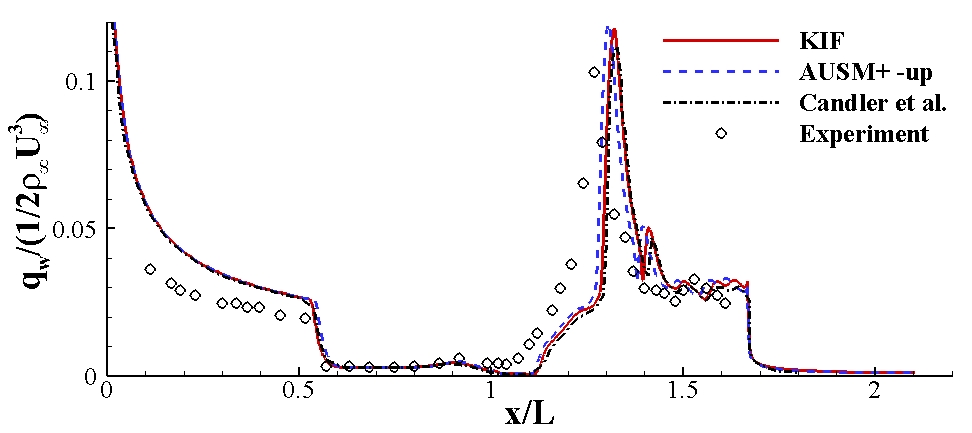}
}
\caption{\label{Fig:hdckif_pq} The curves of the (a) pressure coefficient and (b) heat flux of the hypersonic double-cone flow.}
\end{figure}

When simulating this flow, the CFL number is set 25000 and the iteration number is $1 \times 10^6$. The Ma number contours, the pressure contours and the $\beta$ contours, totally and locally, are shown in Fig.~\ref{Fig:hdckif_map}. In Fig.~\ref{Fig:hdckif_ma3}, the fluid field structure is labeled, and Fig.~\ref{Fig:hdckif_ratiof2} shows how free transport mechanism is introduced in this case. The pressure coefficient and heat flux curves at the surface are shown in Fig.~\ref{Fig:hdckif_pq}, with experiment data~\cite{hdcexp,hdcexp2} and the results of Candler et al.~\cite{candler} for comparison. The results of AUSM+ -up using the same mesh is also shown for comparison. The $L$ in Fig.~\ref{Fig:hdckif_pq} is the length of the first cone. About the pressure coefficient and heat flux curves at the surface, the results of the KIF match well with the results of AUSM+ -up and the results of Candler et al. However, there are errors between the numerical results and the experiment data, about the value of the heat flux peak and the position where the peaks of pressure coefficient and heat flux exist. The reason may be that the thermochemical state doesn't fully de-excite in the experiment. Discussion in detail about this phenomenon can be referred to Ref.~\cite{babinsky,hdcref}.

\subsection{The hypersonic double-ellipsoid flow}
\begin{figure}[!htp]
\centering
\includegraphics[width=0.5\textwidth]{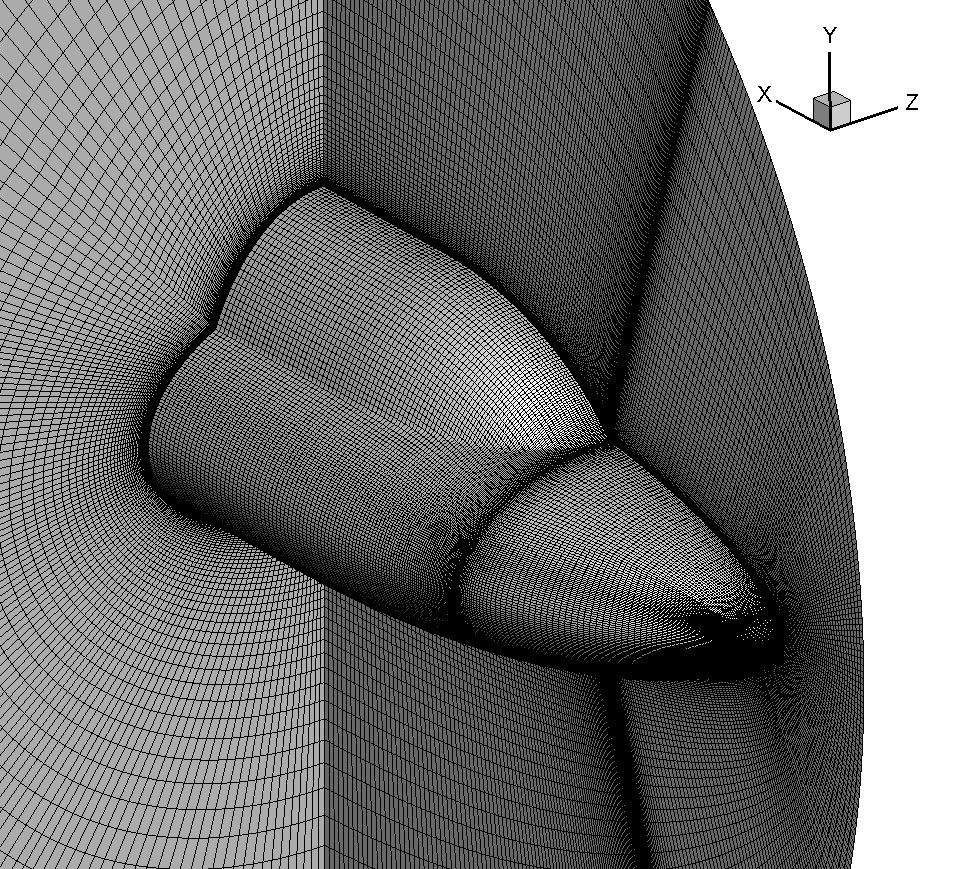}
\caption{\label{Fig:demesh} The mesh of the hypersonic double-ellipsoid flow.}
\end{figure}

This test case is a complex three dimensional hypersonic test case. A hypersonic flow passes the first ellipsoid and forms a bowl shock, intersecting with the other shock generated from the second ellipsoid. The inflow $Ma$ number is 10.02 and the $Re$ number is $2.2 \times 10^6/m$. The length of the model is 0.215m. The inflow temperature is 69K. On the model, the isothermal boundary condition is chosen and the temperature of the wall is 288K. The attack angle is $0^{\circ}$. The mesh is shown in Fig.~\ref{Fig:demesh}. The element number of the mesh is 2210954. The first layer adjacent to the model is $2.0\times10^{-6}m$.

\begin{figure}[!htp]
\centering
\includegraphics[width=0.5\textwidth]{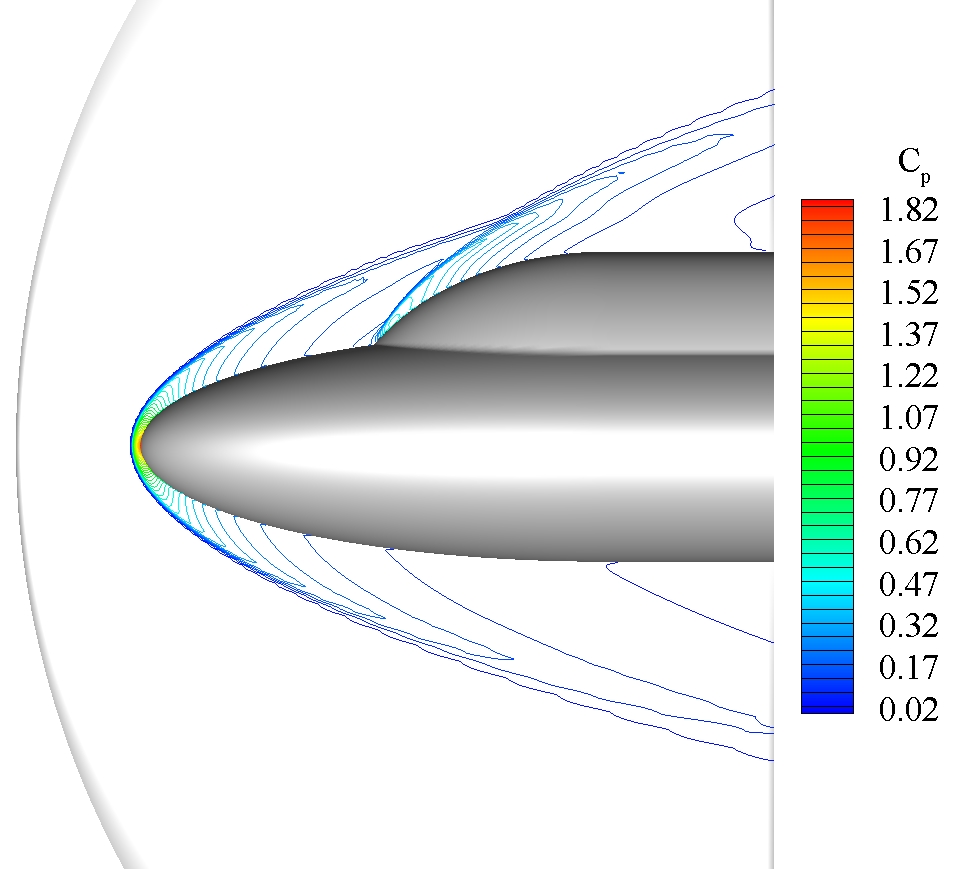}
\caption{\label{Fig:decpd3} The pressure coefficient contour at the symmetry plane of the hypersonic double-ellipsoid flow.}
\end{figure}

\begin{figure}[!htp]
\centering
\subfigure[]{
\includegraphics[width=0.45\textwidth]{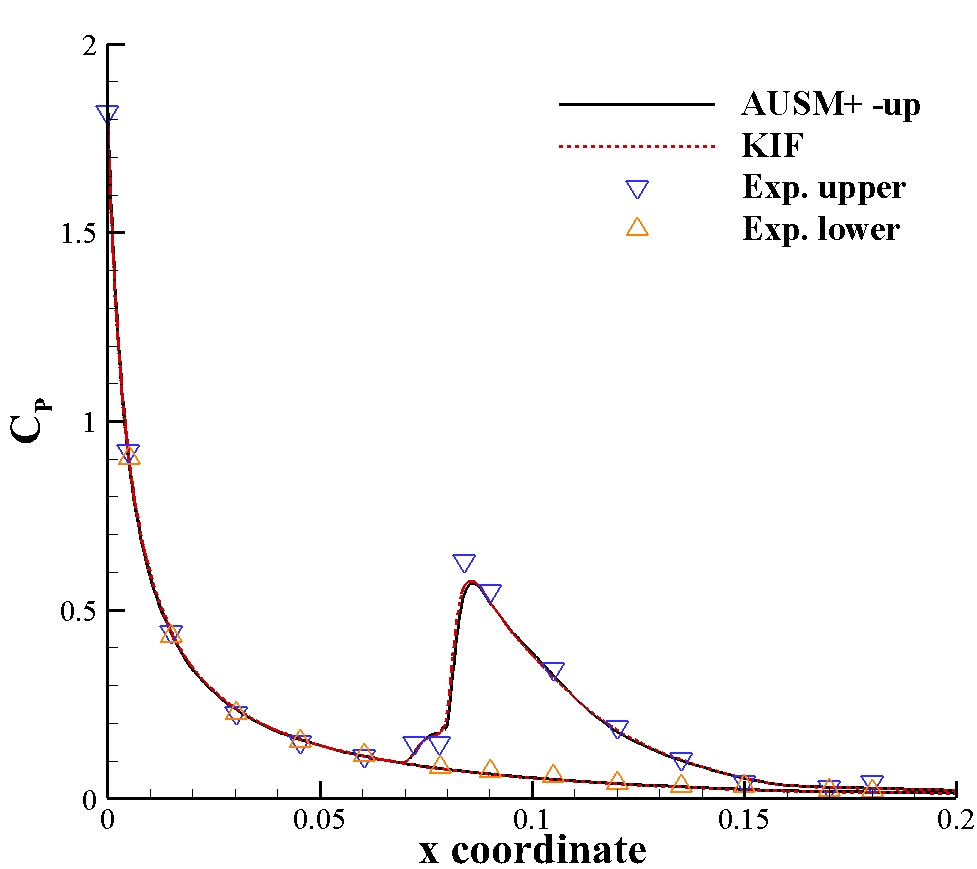}
}
\subfigure[]{
\includegraphics[width=0.45\textwidth]{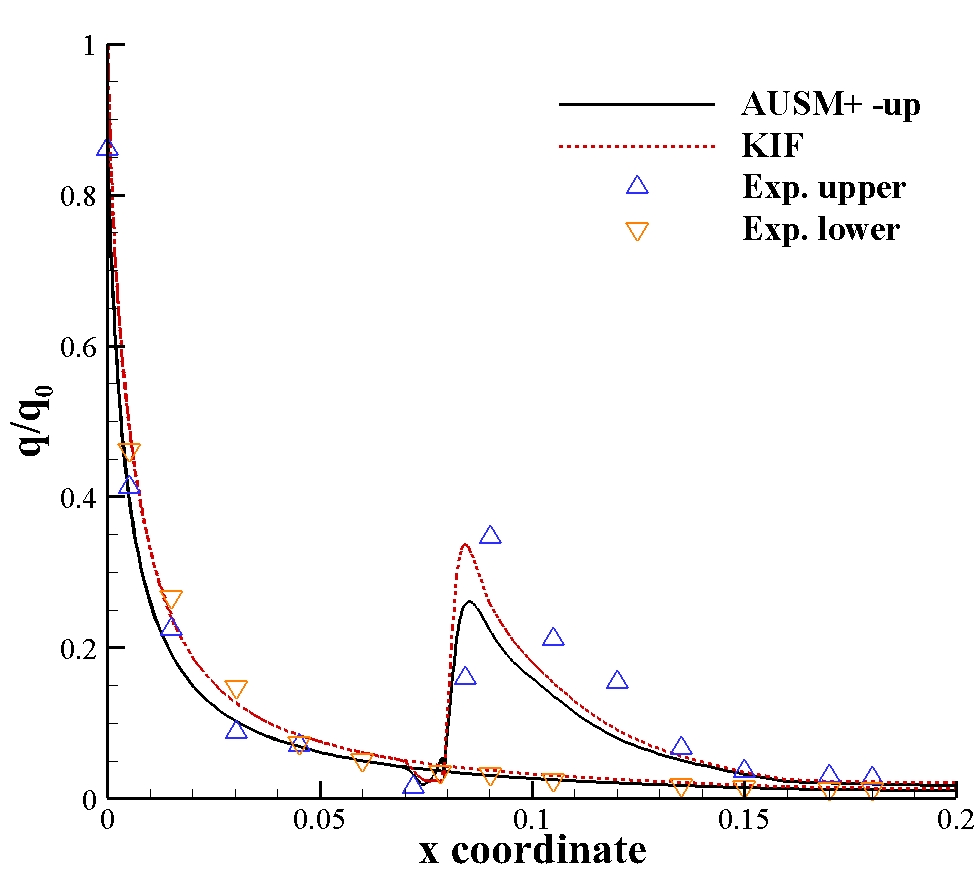}
}
\caption{\label{Fig:depq} The curves of the (a) pressure coefficient and (b) heat flux on the wall at the symmetry plane of the hypersonic double-ellipsoid flow.}
\end{figure}

The SST turbulence model is chosen in this case. At the beginning, the residual is $10^{-7.5}$, and the CFL number is set 200. After $1.2 \times 10^5$ iterations, the residual is $10^{-10.5}$, and the CFL number increases to 4000. The pressure coefficient contour at the symmetry plane is shown in Fig.~\ref{Fig:decpd3}. The pressure coefficient curve and heat flux curve at the wall of the symmetry plane are shown in Fig.~\ref{Fig:depq}. The results of AUSM+ -up and the experiment data~\cite{expde} are used for comparison. The results of the KIF method agree well with both of them. The present method behaves well against the shock instability phenomenon, which means that it is good at solving external flows like this case. As a result, in Fig.~\ref{Fig:depq}, the heat flux curve predicted by the KIF is closer to the experiment data when compared with the result of AUSM+ -up.

\section{\label{sec:conclusion}Conclusion}
In this paper, a modified KIF is developed and an implicit strategy is designed. Six numerical test cases, about continuum flows, covering the whole speed regime, are conducted to examine the validity and the accuracy of the present method. Efficiency is tested in the simulations of the lid-driven cavity subsonic flow and the hypersonic viscous flow past a cylinder. The iteration number of the implicit KIF is always less than 1/800 of the explicit one, validating good efficiency in both subsonic flows and hypersonic flows. Mechanisms are noticed that the KFVS part of the KIF in boundary layers not only imports error but also reduces the CFL number seriously, and that the lacking of KFVS around shock waves reduces the robustness and the accuracy. Corresponding technologies are proposed, through the using of Ma and $\Delta P$, and smoothness is protected by a designed function $F_S(x,a,b)$. Good results are obtained when simulating typical supersonic/hypersonic flows and complex hypersonic flows. Pressure, friction coefficient and heat flux can be simulated accurately in test cases including the hypersonic viscous flow past a cylinder, the laminar shock boundary layer interaction, the supersonic flow around a ramp segment, the hypersonic double-cone flow and the hypersonic double-ellipsoid flow, compared with experiment data and results of other numerical schemes like the AUSM+ -up. Because the free transport mechanism is enough around shock waves, the scheme behaves well against the shock instability phenomenon, which means it is good at solving external flows like the hypersonic viscous flow past a cylinder and the hypersonic double-ellipsoid flow.

\begin{acknowledgments}
The authors thank Prof. Kun Ye at Northwestern Polytechnical University for mesh of the hypersonic double-ellipsoid flow. Junzhe Cao thanks Dr. Ji Li at Northwestern Polytechnical University for helps about designing the implicit strategy. We imbed the present implicit KIF method in the Stanford University Unstructured (SU2) open-source platform~\cite{su1,su3}, and the numerical study is conducted on this platform. We appreciate the SU2 team for their great work. The present work is supported by the National Numerical Wind-Tunnel Project of China and National Natural Science Foundation of China (Grants No. 11702223, No. 11902266, and No. 11902264).
\end{acknowledgments}

\section*{DATA AVAILABILITY}
The data that support the findings of this study are available from the corresponding author upon reasonable request.

\bibliographystyle{unsrt}
\bibliography{lusgs}
\clearpage

\begin{table}
\centering
\caption{\label{table:vertex} The vortex centers of cavity flow at $Re = 1000$.}
\begin{tabular}{p{0.2\columnwidth}<{\centering} p{0.12\columnwidth}<{\centering} p{0.12\columnwidth}<{\centering} p{0.12\columnwidth}<{\centering} p{0.12\columnwidth}<{\centering} p{0.12\columnwidth}<{\centering} p{0.12\columnwidth}<{\centering}}
    \bottomrule[1.4pt]
    \hline
    \toprule
    \hline
    \hline
    \multirow{2}{*}{Method}
    &\multicolumn{2}{p{0.24\columnwidth}<{\centering}}{Central vertex} &\multicolumn{2}{p{0.24\columnwidth}<{\centering}}{Bottom-left vertex} &\multicolumn{2}{p{0.24\columnwidth}<{\centering}}{Bottom-right vertex}\\
    \cline{2-3} \cline{4-5} \cline{6-7}
    &x &y &x &y &x &y\\
    \hline
    Ghia et al.~\cite{ghia}	&0.5313 &0.5625	&0.0859	&0.0781	&0.8594	&0.1094\\
    \hline
    The explicit KIF	&0.5321	&0.5656	&0.0835	&0.0768	&0.8676	&0.1116\\
    \hline
    The implicit KIF	&0.5320	&0.5656	&0.0835	&0.0768	&0.8675	&0.1116\\
    \bottomrule[1.4pt]
    \hline
    \toprule
    \hline
    \hline
\end{tabular}
\end{table}

\end{document}